\documentclass[a4paper,12pt,fleqn]{article}
\usepackage{amsmath,bm}
\usepackage{amssymb}
\usepackage{theorem}
\usepackage{mathrsfs} 
\usepackage[driverfallback=dvipdfm]{hyperref}
\usepackage[usenames,dvipsnames]{pstricks}
\usepackage{euscript}
\usepackage{graphicx}
\usepackage{enumitem}
\usepackage{charter}
\usepackage{empheq}

\graphicspath{{figures/}}
\usepackage{subfigure}

\newcommand{\email}[1]{\href{mailto:#1}{\nolinkurl{#1}}}
\definecolor{labelkey}{rgb}{0,0.08,0.45}
\definecolor{refkey}{rgb}{0,0.6,0.0}
\definecolor{Brown}{rgb}{0.45,0.0,0.05}
\definecolor{dgreen}{rgb}{0.00,0.49,0.00}
\definecolor{dblue}{rgb}{0,0.08,0.75}
\hypersetup{linktocpage=true,citecolor=dblue}

\PassOptionsToPackage{normalem}{ulem}
\usepackage{ulem}

\numberwithin{equation}{section}
\newcommand{\bb}{\bm{b}}
\newcommand{\dd}{\bm{d}}
\newcommand{\DD}{\ensuremath{{\EuScript D}}}
\newcommand{\LL}{\ensuremath{\mathbb{\mathcal{L}}}}
\newcommand{\ff}{\bm{f}}
\newcommand{\WW}{\bm{W}}
\newcommand{\FF}{\ensuremath{{\mathcal F}}}
\newcommand{\uu}{\bm{u}}
\newcommand{\TT}{\ensuremath{\mathcal{T}}}
\newcommand{\sign}{\ensuremath{\text{\rm sign}}}
\newcommand{\yy}{\bm{y}}
\newcommand{\Wu}{\bm{W}\bm{u}}
\newcommand{\XX}{\bm{X}}
\newcommand{\YY}{\bm{Y}}
\newcommand{\ZZ}{\bm{Z}}
\newcommand{\calW}{{\ensuremath{\boldsymbol{\mathcal W}}}}
\newcommand{\ww}{\bm{w}}
\newcommand{\calN}{{\ensuremath{\boldsymbol{\mathcal N}}}}
\usepackage{mathptmx} 

\begin{document}
\title{\sffamily \vskip -9mm
A Review on Deep Learning in Medical Image Reconstruction \thanks{The work of Haimiao Zhang is funded by China Postdoctoral Science Foundation under grant No. 2018M641056. The work of Bin Dong is supported in part by  National Natural Science Foundation of China (NSFC) grant No. 11831002, and Beijing Natural Science Foundation (No. Z180001).\protect\\
\small $\!^\dagger$Beijing International Center for Mathematical Research (BICMR), Peking University, Beijing, 100871, P. R. China.
Email:
\email{hmzhang@pku.edu.cn, dongbin@math.pku.edu.cn}
\protect\\
\small $\!^\ddagger$Contact author:
Bin Dong, \email{dongbin@math.pku.edu.cn}, phone: +86 10-62744091.
}}
\author{
Haimiao Zhang$^\dagger$ and
Bin Dong$^\dagger$ $^\ddagger$ 
}
\maketitle
\thispagestyle{empty}

\vskip -5mm

\noindent
{\bfseries Abstract.}
Medical imaging is crucial in modern clinics to provide guidance to the diagnosis and treatment of diseases. Medical image reconstruction is one of the most fundamental and important components of medical imaging, whose major objective is to acquire high-quality medical images for clinical usage at the minimal cost and risk to the patients. Mathematical models in medical image reconstruction or, more generally, image restoration in computer vision, have been playing a prominent role. Earlier mathematical models are mostly designed by human knowledge or hypothesis on the image to be reconstructed, and we shall call these models handcrafted models. Later, handcrafted plus data-driven modeling started to emerge which still mostly relies on human designs, while part of the model is learned from the observed data. More recently, as more data and computation resources are made available, deep learning based models (or deep models) pushed the data-driven modeling to the extreme where the models are mostly based on learning with minimal human designs. Both handcrafted and data-driven modeling have their own advantages and disadvantages. Typical handcrafted models are well interpretable with solid theoretical supports on the robustness, recoverability, complexity, etc., whereas they may not be flexible and sophisticated enough to fully leverage large data sets. Data-driven models, especially deep models, on the other hand, are generally much more flexible and effective in extracting useful information from large data sets, while they are currently still in lack of theoretical foundations. Therefore, one of the major research trends in medical imaging is to combine handcrafted modeling with deep modeling so that we can enjoy benefits from both approaches. The major part of this article is to provide a conceptual review of some recent works on deep modeling from the unrolling dynamics viewpoint. This viewpoint stimulates new designs of neural network architectures with inspirations from optimization algorithms and numerical differential equations. Given the popularity of deep modeling, there are still vast remaining challenges in the field, as well as opportunities which we shall discuss at the end of this article.
{\bfseries Subclass.} 60H10, 92C55,  93C15, 94A08

\section{Introduction}
\label{intro}
Medical image reconstruction can often be formulated as the following mathematical problem
\begin{equation}\label{eq:general-imaging-model}
\bm{f}=\bm{A}\bm{u}\oplus \bm{\eta},
\end{equation}
where $\bm{A}$ is a physical system modeling the image acquisition process. Operator $\bm{A}$ can be a linear operator or nonlinear operator that depends on the specific imaging modality. Variable $\bm{u}$ is the unknown image to be reconstructed, and $\bm{f}$ is the measured data that might be contaminated by noise $\bm{\eta}$ with known or partially known noise statistics, e.g., Gaussian, Laplacian, Poisson, Rayleigh, etc. The operator $\oplus$ is a notation to denote addition when Gaussian noise is assumed, a certain nonlinear operator when Poisson noise or Rician noise is assumed. In different image reconstruction tasks, $\bm{A}$ takes different forms:
\begin{itemize}
\item Denoising:
$\bm{A}$ is an identity operator.
\item Deblurring:
$\bm{A}$ is a convolution operator. When the convolution kernel is unknown, the problem is called blind deblurring \cite{pavlovic1992maximum}.
\item Inpainting:
$\bm{A}$ is a restriction operator which can be represented by a diagonal matrix with value 0 or 1 \cite{BSCB}.
\item Magnetic resonance imaging (MRI):
$\bm{A}$ is a sub-sampled Fourier transform which is a composition of the Fourier transform and a binary sampling operator \cite{brown2014magnetic}.
\item X-Ray based computed tomography (CT):
$\bm{A}$ is a sub-sampled Radon transform, which is a partial collections of line integrations \cite{buzug2008computed}.
\item Quantitative susceptibility mapping (QSM) \cite{choi2014inverse,natterer2016image,de2010quantitative,wang2015quantitative}: $\bm{A}$ is the dipole kernel $$A(X)=\frac{2z^2-x^2-y^2}{4\pi (x^2+y^2+z^2)^{5/2}},\ X=(x,y,z)\in \mathbb{R}^{3}.$$
\end{itemize}
The inverse problem \eqref{eq:general-imaging-model} is in general challenging to solve due to the large-scale and ill-posed nature of the problem in practice.

\subsection{Image reconstruction models}\label{sub:image_reconstruction_models}

The above inverse problem \eqref{eq:general-imaging-model} covers a wide range of image restoration tasks which are not limited to medical image reconstruction. To solve the inverse problem \eqref{eq:general-imaging-model}, it is common practice to consider the following optimization problem
\begin{equation}\label{eq:optimization-model}
\min_{\bm{u} \in \DD} \LL(\bm{u})=F(\bm{A}\bm{u},\ff)+\lambda \Phi(\WW,\bm{u}).
\end{equation}
The solution $\bm{u}^\star=\arg\min_{\bm{u}}\LL(\bm{u})$ is an approximate solution to the inverse problem \eqref{eq:general-imaging-model}. In \eqref{eq:optimization-model}, the term $F(\bm{A}\bm{u},\ff)$ is the data fidelity term that measures the consistency of the approximate solution to the measured data $\ff$. Its specific form normally depends on the noise statistics. For example:
\begin{itemize}
\item Gaussian noise:
$F(\bm{A}\bm{u},\ff)=\frac{1}{2}\|\bm{A}\bm{u}-\ff\|_{2}^{2}$,
\item Poisson noise:
$F(\bm{A}\bm{u},\ff)=\langle \bm{1}, \bm{A}\bm{u}\rangle -\langle \ff, \log(\bm{A}\bm{u})\rangle$,
with $\langle \bm{a},\bm{b}\rangle=\sum_{i} \bm{a}_{i}\bm{b}_{i}$,
\item Impulse noise:
$F(\bm{A}\bm{u},\ff)=\|\bm{Au}-\bm{f}\|_{1}$,
\item Multiplicative noise \cite{Rudin2003Multiplicative,dong2013convex}:
$F(\bm{A}\bm{u},\ff)=\langle \log(\bm{Au}) +\frac{\bm{f}}{\bm{Au}}, \bm{1}\rangle +\lambda\|\sqrt{\frac{\bm{Au}}{\bm{f}}}-\bm{1}\|^{2}.$
\end{itemize}
The second term $\Phi(\WW,\bm{u})$ in \eqref{eq:optimization-model} is the regularization term encoding the prior knowledge on the image to be reconstructed. The regularization term is often the most crucial part of the modeling, and what people have mostly focused on in the literature. The parameter $\lambda$ in \eqref{eq:optimization-model} provides a balance between the data fidelity term and the regularization term. Mathematical modeling has been playing a vital role in solving such inverse problems. Interested reader can refer to  \cite{aubert2006mathematical,ChanShen,dong2015image,shen2010wavelet} for more extensive reviews on mathematical models for image restoration.

Deep learning models can also be casted into the form of \eqref{eq:optimization-model}. However, there are differences as well. In handcraft or handcraft $+$ data-driven modeling, the transformation $\WW$ is often a certain linear transformation that is able to extract sparse features from the images. In handcraft models, $\WW$ is often given by design (e.g. a differential operator or wavelet transform); in handcraft $+$ data-driven models, $\WW$ (or a portion of it) is often learned from the given data. Sparsity is the key to the success of these models. Deep learning models follow a similar modeling philosophy by considering nonlinear sparsifying transformations rather than linear ones. In general, we define a parameterized nonlinear mapping $\mathcal{V}(\cdot,\bm{\Theta}):\FF\to \mathcal{U}, \ff \mapsto \uu$ that maps the input data $\ff$ to a high quality output image $\bm{u}$. The mapping $\mathcal{V}$ is parameterized by $\bm{\Theta}$ which is trained on a given data set $\FF \times\mathcal{U}$ by solving the following optimization problem
$$\min_{\Theta}\frac{1}{\#(\FF \times\mathcal{U})}\sum_{(\ff,\bm{u})\in \FF \times\mathcal{U}}\mathcal{C}(\mathcal{V}(\ff,\bm{\Theta}),\bm{u}),$$
where $\mathcal{C}(\cdot,\cdot)$ is a metric of difference between the approximated image $ \mathcal{V}(\ff,\bm{\Theta})$ and the ground truth image $\bm{u}$, and $\#(\FF \times\mathcal{U}) $ is the cardinality of the data set $\FF \times\mathcal{U}$.
To prevent overfitting, we can introduce a regularization term to the above optimization problem as in \eqref{eq:optimization-model}. We then have the problem
\begin{equation}\label{eq:NN-image-process-model}
\min_{\bm{\Theta}}\mathcal{L}(\ff,\uu;\bm{\Theta})=\frac{1}{\#(\FF \times\mathcal{U})}\sum_{(\ff,\bm{u})\in \FF \times\mathcal{U}}\mathcal{C}(\mathcal{V}(\ff,\bm{\Theta}),\bm{u}) +\mathcal{R}(\bm{\Theta}),
\end{equation}
where $\mathcal{R}(\cdot)$ is the regularization term that can be chosen as, for example, the $\ell_{2}$ or $\ell_{1}$ norm. Good examples of the nonlinear mapping $\mathcal{V}(\cdot)$ include the stacked denoising autoencoder (SDAE) \cite{Vincent2010Stacked}, the U-Net \cite{Ronneberger2015U-net}, the ResNet \cite{He2016Deep,He2016Identity}, etc. We postpone a detailed discussion on these networks and how to interpret them in mathematical terms in later sections.

The development of modeling in image reconstruction for the past three decades can be summarized to three stages:
\begin{itemize}
\item
\textbf{Handcraft modeling (1990-).} models are designed based on mathematical characterizations on the desirable recovery of the image. For example, the ideal function space a ``good" image should belong to, ideal local or global geometric properties the image should enjoy, or sparse approximation by certain well-designed basis functions, etc. Successful handcraft models include total variation (TV) \cite{rudin1992nonlinear} model, Perona-Malik diffusion \cite{perona1994anisotropic,perona1990scale}, shock-filters  \cite{osher1990feature,alvarez1994signal}, nonlocal methods \cite{buades2005non,buades2005review,buades2010image,lou2010image,dabov2007image}, wavelet  \cite{daubechies1992ten,Mallat2009}, wavelet frames  \cite{ron1997affine,Dong2010IASNotes}, BM3D \cite{dabov2007image}, WNNM \cite{gu2014weighted}, etc. These models mostly have solid theoretical foundations and high interpretability. They work reasonably well in practice, and some of them are still the state-of-the-art methods for certain tasks.

\item \textbf{Handcraft $+$ data-driven modeling (1999-).}
Starting from around 1999, models that combine data-driven or learning with handcraft modeling started to emerge. These models rely on some general mathematical or statistical framework by handcraft designs. However, the specific form of the model is determined by the given image data or data set. Comparing to purely handcrafted models, these models can better exploit the available data and outperform their corresponding none data-driven counterparts. Meanwhile, the handcrafted framework of the models grants certain interpretability and theoretical foundation to the models. Successful examples include the method of optimal directions \cite{Engan1999Method}, the K-SVD \cite{aharon2006K-SVD}, learning based PDE design \cite{Liu2010Learning}, data-driven tight frame \cite{cai2014data,bao2015convergence}, Ada-frame \cite{tai2016multiscale},
low-rank models \cite{wright2009robust,liu2013robust,Jian2014Cine,candes2009exact,cai2010singular}, piecewise-smooth image models \cite{mumford1989optimal,cai2016image}, and statistical models \cite{heimann2009statistical}, etc.

\item \textbf{Deep learning models (2012-).} 2012 is the year that signifies the uprise of deep learning in computer vision with the introduction of a convolutional neural network (CNN) called  AlexNet \cite{Krizhevsky2012ImageNet} for image classification. Then, various types of CNNs such as ResNet \cite{He2016Deep,He2016Identity} and generative adversarial networks (GANs) \cite{goodfellow2014generative} were introduced and applied in image reconstructions. We shall refer to these models as deep learning based models (or deep models for short). Most deep models have millions to billions of parameters. These parameters are trained (optimized) on large data sets via parallel computing (e.g., on graphics processing units (GPUs)). Deep models have greatly advanced the state-of-the-art of many image reconstruction tasks and have changed the research landscape of computer vision in general. The success of deep models is mainly due to the presence of large image data sets with high-quality labels, and the accessibility of massive computing resources. The reliance of deep models on large labeled data sets limits, at least for the moment, the application of deep learning in medical image reconstruction or healthcare in general. The major focus of this review is to recall and discuss deep models in medical image reconstruction, and the limitations, challenges, and opportunities in this new and exciting research direction.
\end{itemize}

Note that what makes medical image reconstruction different from image restoration in computer vision is quality metrics on the reconstructed image. Although researchers use standard metrics such as the peak signal to noise ratio (PSNR), mean square error, structure similarity (SSIM), etc., meaningful quality metrics of a reconstructed medical image should be clinically relevant and task dependent. Furthermore, most medical images are 3D arrays which poses computation challenge as well.

\subsection{Algorithm design for image reconstruction models}\label{sub:algorithm_design_for_image_reconstruction_models}

The difficulties of solving the image reconstruction models motivate the optimization community to design highly efficient numerical algorithms for large scale, nonsmooth and even nonconvex optimization problems. Representative algorithms include the alternating direction method of multipliers (ADMM) \cite{boyd2011distributed,Gabay1976A,Glowinski1975Sur}, primal-dual algorithm \cite{zhu2008efficient,esser2010general,chambolle2011first}, split Bregman algorithm \cite{cai2009split,goldstein2009split}, linearized Bregman algorithm \cite{yin2008bregman,osher2010fast} iterative shrinkage-thresholding algorithm (ISTA) \cite{daubechies2004iterative}, and fast iterative shrinkage-thresholding algorithm (FISTA) \cite{beck2009fast}, among many others. Here, we briefly review some of the algorithms that will be needed in later sections.

\subsubsection{ISTA and FISTA}

Consider the following optimization problem which is a special case of \eqref{eq:optimization-model}
\begin{equation}\label{eq:ISTA-optim}
\min_{\bm{\alpha}} \frac{1}{2}\|\ff-\WW^{\top}\bm{\alpha}\|_2^{2}+\lambda \|\bm{\alpha}\|_{1},
\end{equation}
where $\WW^{\top} $ is a decoding operator that maps code $\bm\alpha$ back to image domain.
Then, the iterative soft-thresholding algorithm (ISTA) solving \eqref{eq:ISTA-optim} simply reads as
\begin{equation}\label{eq:ISTA-solution}
\bm{\alpha}^{k+1}=\TT_{\lambda\tau_{k}}\left(\bm{\alpha}^{k}-2\tau_{k}\WW(\WW^{\top}\bm{\alpha}^{k}-\ff)\right),
\end{equation}
where $ \tau_{k}>0$ is an appropriate step size and the soft-thresholding operator $\TT_{\lambda}(\cdot)$ is defined component-wisely as $\TT_{\lambda}(x)=\sign(x)\max(|x|-\lambda,0),\mbox{with} \; x\in \mathbb{R}.$ ISTA was explicitly proposed in \cite{daubechies2004iterative}. Its idea, however, can be traced back to the classical proximal forward-backward algorithm \cite{bruck1977weak,passty1979ergodic}. Later, an accelerated version of ISTA, called fast iterative soft-thresholding algorithm (FISTA) was introduced \cite{beck2009fast,Shen2011An} which is based on the idea of Nesterov's \cite{nesterov1983method}. FISTA takes following form
\begin{eqnarray}\label{eq:FISTA-scheme}
\bm{\alpha}^{k+1}&=&\TT_{\lambda/L_{\rm{lip}}}\left(\yy^{k}-\frac{1}{L_{\rm{lip}}}\WW(\WW^{\top}\bm{y}^{k}-\ff)\right),\notag\\
t_{k+1}&=& \frac{1+\sqrt{1+4t_{k}^{2}}}{2},\notag\\
\yy^{k+1}&=& \bm{\alpha}^{k+1}+\frac{t_{k}-1}{t_{k+1}}(\bm{\alpha}^{k+1}-\bm{\alpha}^{k}),
\end{eqnarray}
where $L_{\rm{lip}}$ is the Lipschitz constant of the quadratic term in \eqref{eq:ISTA-optim}.

\subsubsection{ADMM/Split Bregman Algorithm}\label{ssub:admm_algorithm}

Consider the following special case of the optimization problem \eqref{eq:optimization-model}
$$\min_{\uu} \LL(\bm{u})=\frac{1}{2}\|\bm{A}\bm{u}-\ff\|_2^{2}+\lambda \|\WW\bm{u}\|_1,$$
which can be written equivalently as
$$\min_{\uu,\dd} \LL(\uu,\dd)=\frac{1}{2}\|\bm{A}\bm{u}-\ff\|_2^{2}+\lambda \|\dd\|_1,\quad s.t. \quad \WW\uu=\dd.$$
The corresponding augmented Lagrangian function \cite[Chapter 17]{Nocedal2006Numerical} is defined by
$$ \LL(\uu,\dd;\bb)=\frac{1}{2}\|\bm{A}\bm{u}-\ff\|_2^{2}+\lambda \|\dd\|_1+\langle \WW\uu-\dd,\bb\rangle +\frac{\mu}{2}\|\Wu-\dd\|_2^{2}, $$
with the Lagrangian multiplier $\bb$. Then, the alternating direction method of multipliers (ADMM) or split Bregman algorithm takes the form \cite{goldstein2009split,boyd2011distributed}
\begin{eqnarray}\label{eq:ADMM-scaled-form}
\uu^{k+1}&=&\left(\bm{A}^{\top}\bm{A}+\mu \WW^{\top}\WW\right)^{-1}\left[\bm{A}^{\top}\ff+\mu \WW^{\top}(\dd^{k}-\bm{\nu}^{k})\right],\notag\\
\dd^{k+1}&=&\TT_{\lambda/\mu}\left(\Wu^{k+1}+\bm{\nu}^{k}\right),\notag\\
\bm{\nu}^{k+1}&=& \bm{\nu}^{k}+(\Wu^{k+1}-\dd^{k+1}),
\end{eqnarray}
where $\mu$ is a tuning parameter.

\subsubsection{The Primal-Dual Algorithm}\label{ssub:primal_dual_algorithm}

Consider the following optimization problem
\begin{equation}\label{eq:primal-optim-prob}
\min_{\uu} F(\bm{u})+ \Phi(\Wu),
\end{equation}
where $ F(\bm{u})$ is the data fidelity term and $\Phi(\Wu) $ is the regularization term appeared in \eqref{eq:optimization-model}. Assume $F:\mathbb{R}^{n}\to (-\infty,+\infty]$ and $\Phi:\mathbb{R}^{m}\to (-\infty,+\infty]$ are closed  proper convex functions. The problem \eqref{eq:primal-optim-prob} can be written equivalently as
\begin{equation}\label{eq:PD-optim-prob}
\min_{\uu}\max_{\bm{w}} F(\bm{u})+ \langle \Wu,\bm{w}\rangle -\Phi^{\ast}(\bm{w}),
\end{equation}
Then, the primal-dual hybrid gradient (PDHG) algorithm \cite{zhu2008efficient,esser2010general,chambolle2011first} can be written as
\begin{eqnarray}\label{eq:PD-resolvent}
\bm{w}^{k+1}&=&(I+\partial\Phi^{\ast})^{-1}\left( \bm{w}^{k}+\alpha_{k}\Wu^{k}\right) ,\notag\\
\uu^{k+1}&=&(I+\partial F)^{-1}\left( \uu^{k}-\beta_{k}\WW^{\top}\bm{w}^{k+1}\right),
\end{eqnarray}
where $\alpha_{k}$ and $\beta_{k}$ are tuning parameters. Note that in \cite{chambolle2011first}, the authors introduced an additional correction update step,
\begin{equation}\label{eq:PD-error-correct}
\bar{\uu}^{k+1}=\uu^{k+1}+\theta(\uu^{k+1}-\uu^{k}),
\end{equation}
to the original PDHG algorithm \eqref{eq:PD-resolvent} and replaced $\uu^{k}$ in $\bm{w}^{k+1}$-step by $ \bar{\uu}^{k}$.

\subsubsection{SGD}

It is very common in machine learning that an optimization problem takes the following form
\begin{equation}\label{eq:ML-optim}
\min_{\bm{\Theta}}F_{N}(\bm{\Theta})=\frac{1}{N}\sum_{i=1}^{N}f_{i}(\bm{\Theta}).
\end{equation}
The main computation challenge, especially in deep learning, is that $N$ can be huge, e.g., in the magnitude of millions to billions. Therefore, the evaluation of the function value $F_N$ and its gradient can be rather slow. In such case, stochastic gradient descent (SGD) algorithm \cite{Bottou2010Large,robbins1951stochastic,bottou2012stochastic,zhang2004solving} and its variants \cite{nitanda2014stochastic,zhang2017stochastic,konevcny2016mini} are among the most popular algorithms in deep learning.

The very basic form of (mini-batch) SGD is
$$\bm{\Theta}^{k+1}=\bm{\Theta}^{k}-\alpha_{k}\frac{1}{|\mathcal{S}_{k}|}\sum_{i_{k}\in\mathcal{S}_{k}}\nabla f_{i_{k}}(\bm{\Theta}^{k}),$$
where $\alpha_{k}$ is the step size (or learning rate) and $\mathcal{S}_{k}$ is a random subset of $\{1,2,\ldots,N\}$. The evaluation of $\frac{1}{|\mathcal{S}_{k}|}\sum_{i_{k}\in\mathcal{S}_{k}}\nabla f_{i_{k}}(\bm{\Theta}^{k})$ provides an unbiased estimation of the full gradient and is computationally cheap. Other than SGD, numerous randomized algorithms are being used in deep learning, such as Adam \cite{kingma2014adam}, AdaGrad \cite{duchi2011adaptive}, RMSProp \cite{hinton2012video}, etc. A comprehensive review on optimization algorithms for large scale machine learning problems can be found in  \cite{bottou2018optimization}.

\subsection{When handcraft modeling meets deep learning}

Both handcrafted models and deep models have their advantages and drawbacks depending on the applications. Most handcrafted models are designed with a solid mathematical foundation and can be very well interpreted. However, handcrafted models are not flexible enough to fully leverage large data sets. Deep models, on the other hand, are generally much more flexible and can better extract useful information from large data sets. However, they are generally more challenging to interpret. For the moment, they are also in lack of theoretical foundations in contrast to handcrafted models. Therefore, there has been an increasing effort in the community to combine handcrafted modeling and deep modeling so that we can enjoy benefits from both approaches.

One of the most popular ways of such combination is the so-called unrolling dynamics approach. It started with the work of \cite{gregor2010learning} where the authors showed that one could unroll the iterative soft-thresholding algorithm (ISTA) \eqref{eq:ISTA-solution} to create a feed-forward network. Then, one can train ISTA in an end-to-end fashion to determine the parameters in ISTA so that they are best suitable for the training data. They called the unrolled dynamics LISTA and demonstrated its advantage over ISTA. This work showed that one could unroll a discrete dynamic system to form a network for end-to-end training. More recently, more and more examples showed that the unrolling dynamics approach seems a good balance between model interpretability and efficacy. This includes unrolling discrete forms of nonlinear diffusions for image restoration \cite{Liu2010Learning,chen2015learning} and unrolling optimization algorithms for medical imaging and inverse problems \cite{sun2016deep,adler2018learned,solomon2018deep,chen2018theoretical,liu2018convergence,pmlr-v95-li18f,zhang2019JRSNet}. The unrolling dynamics approach can often result in deep models that have better interpretability inherited from the original dynamics.

Furthermore, these deep models normally have much fewer trainable parameters than black-box deep neural networks, which are more suitable for learning on relatively small data sets. On the other hand, we may interpret certain classes of deep convolutional networks, such as ResNet, as discrete dynamics, and hence relates deep learning with optimal control \cite{weinan2017proposal,chang2017multi}. Such viewpoint not only provides elegant interpretation of deep neural networks \cite{li2017deep}, but also enable us to design more effective deep networks for various tasks in machine learning \cite{chang2018reversible,lu18beyond,li2018optimization,wang2018enresnet,ruthotto2018deep,tao2018nonlocal,zhang2019you}, computer vision \cite{zhang2018dynamically}, inverse problems \cite{long2018pde,long2018pde2}, and natural language processing \cite{lu2019understanding}. More recently, intriguing relations between deep convolutional networks with multigrid method are addressed \cite{xu2019MgNet} which lead to new interpretations to deep models.

The remainder of this paper is organized as follows. In Section \ref{sec:review_of_deep_neural_networks}, we will review some recently proposed deep neural networks which are popular in medical imaging. Section \ref{sec:interpretation_of_dl_architecture} shows the understanding of deep neural networks from the perspective of representation learning and differential equations. Section \ref{sec:imaging_models} reviews some recently proposed deep modeling for medical imaging, where Section \ref{sub:post_processing} presents some examples of post-processing deep models, Section \ref{sec:deep_model_design} collects some models that are designed by combining handcrafted modeling with deep modeling, and Section \ref{sec:task_driven} reviews task-driven deep models. To conclude, Section \ref{sec:challenge_and_opportunities_in_deep_learning} summarizes the main challenge and opportunities in deep learning based medical imaging.

\section{Review of Deep Neural Networks}\label{sec:review_of_deep_neural_networks}

Deep neural networks (DNNs) are now proven to be powerful tools to represent complex data. The main differences between DNNs and traditional machine learning models are the composite nonlinearity of the DNNs and the end-to-end training, which make DNNs very effectively in extracting features that are most suitable for a given task. In recent years, DNNs are used in various medical imaging tasks, including image reconstruction, segmentation, region-of-interest detection, super-resolution, classification, etc. In this section, we briefly recall some of the DNNs that are widely adopted in medical imaging.

\subsection{ResNet}

\begin{figure}[h]
\centering
\includegraphics[scale=0.4]{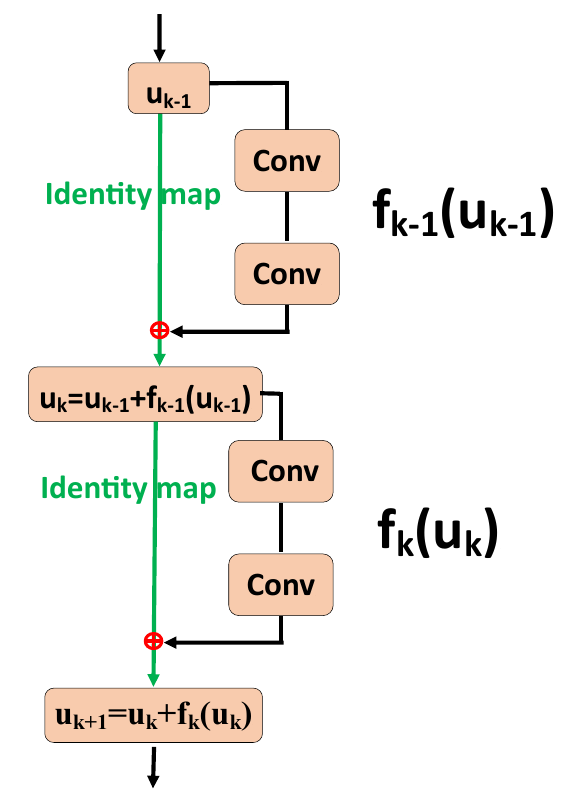}
\caption{ResNet.}\label{fig:ResNet}
\end{figure}

In computer vision, the residual network (ResNet) \cite{He2016Deep,He2016Identity} is one of the most popular DNNs. The architecture of ResNet is shown in Figure \ref{fig:ResNet} which can also be formulated mathematically as
\begin{equation}\label{eq:resnet_block}
\uu_{k+1}=\uu_{k}+\FF_k(\uu_{k}),
\end{equation}
where $\uu_{k}$ (resp. $\uu_{k+1}$) indicates the input (resp. output) feature map of the $k$-th layer of the ResNet and $\FF_k(\uu_{k})$ is called a nonlinear residual block with trainable parameters. The skip connection of ResNet is crucial in facilitating stable training when the network is very deep. Other DNNs with the similar skip connections include the learned diffusion model TRD \cite{chen2015learning}, DenseNet \cite{huang2017densely} and U-Net \cite{Ronneberger2015U-net}, among many others.

\subsection{Autoencoder}\label{sub:auto_encoder}

\begin{figure}[h]
\centering
\includegraphics[width=12.0cm]{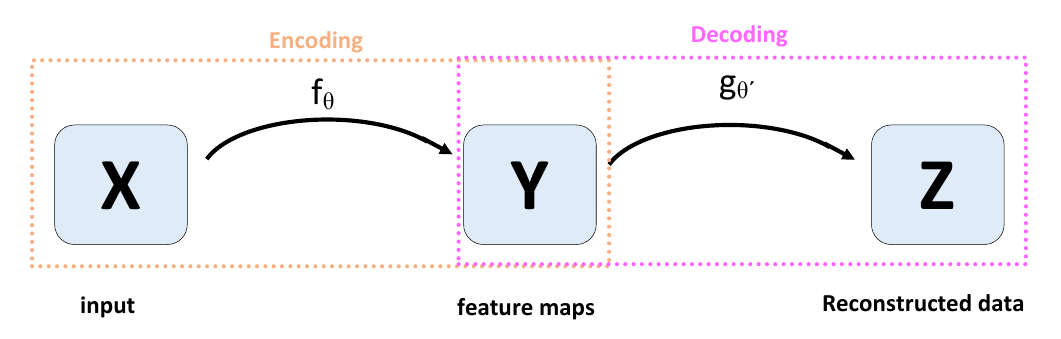}
\caption{Autoencoder.}
\label{fig:autoencoder}
\end{figure}

Autoencoder (AE) \cite{bengio2007greedy,poultney2007efficient} is a type of neural network that is used to learn data representation in an unsupervised manner. It aims to learn an encoder from a set of data together with a decoder so that we do not lose any essential information during the encoding and decoding process. Figure \ref{fig:autoencoder}  presents a typical example of the AE architecture. For a given image $\XX$, the parameterized mapping $f_{\bm{\theta}}$ (e.g. a fully connected or a convolutional neural network) is an encoder that extract feature maps from $\XX$. The encoded multi-channel feature maps are denoted by $\YY=f_{\bm{\theta}}(\XX)$. The encoded feature maps $\YY$ is then decoded by another parameterized mapping  $g_{\bm{\theta}^{\prime}}$ to obtain the reconstructed data $\ZZ$. The parameters $\bm\theta$ and $\bm\theta'$ are optimized on a data set so that a properly chosen loss function that measures the average discrepancies between $\XX$ and $\ZZ$ is minimized. AE resembles linear representations such as Fourier and wavelet transform if we regard encoding as the decomposition, decoding as the reconstruction and feature maps as the coefficients of the representation. However, the representation provided by AE is nonlinear and is learned from a data set.

To learn a more effective and robust representation, \cite{Vincent2010Stacked} proposed the stacked denoising autoencoder (SDAE). In SDAE, the encoder and decoder are DNNs, and they are trained to recover $\ZZ\approx\XX$ from noisy input $\XX$.  Based on the encoder/decoder framework, \cite{Badrinarayanan2017SegNet} designed a DNN, called SegNet, for image segmentation. In \cite{mao2016image}, the encoder/decoder framework is adopted for image denoising and super-resolution. More recently, \cite{chen2017low} designed a residual encoder-decoder CNN to suppress the noise and preserve features in low-dose CT images that are reconstructed using the filtered back projection (FBP) algorithm.

\subsection{U-Net}\label{sub:multi_resolution_analysis_perspective}

\begin{figure}[h]
\centering
\includegraphics[width=12.0cm]{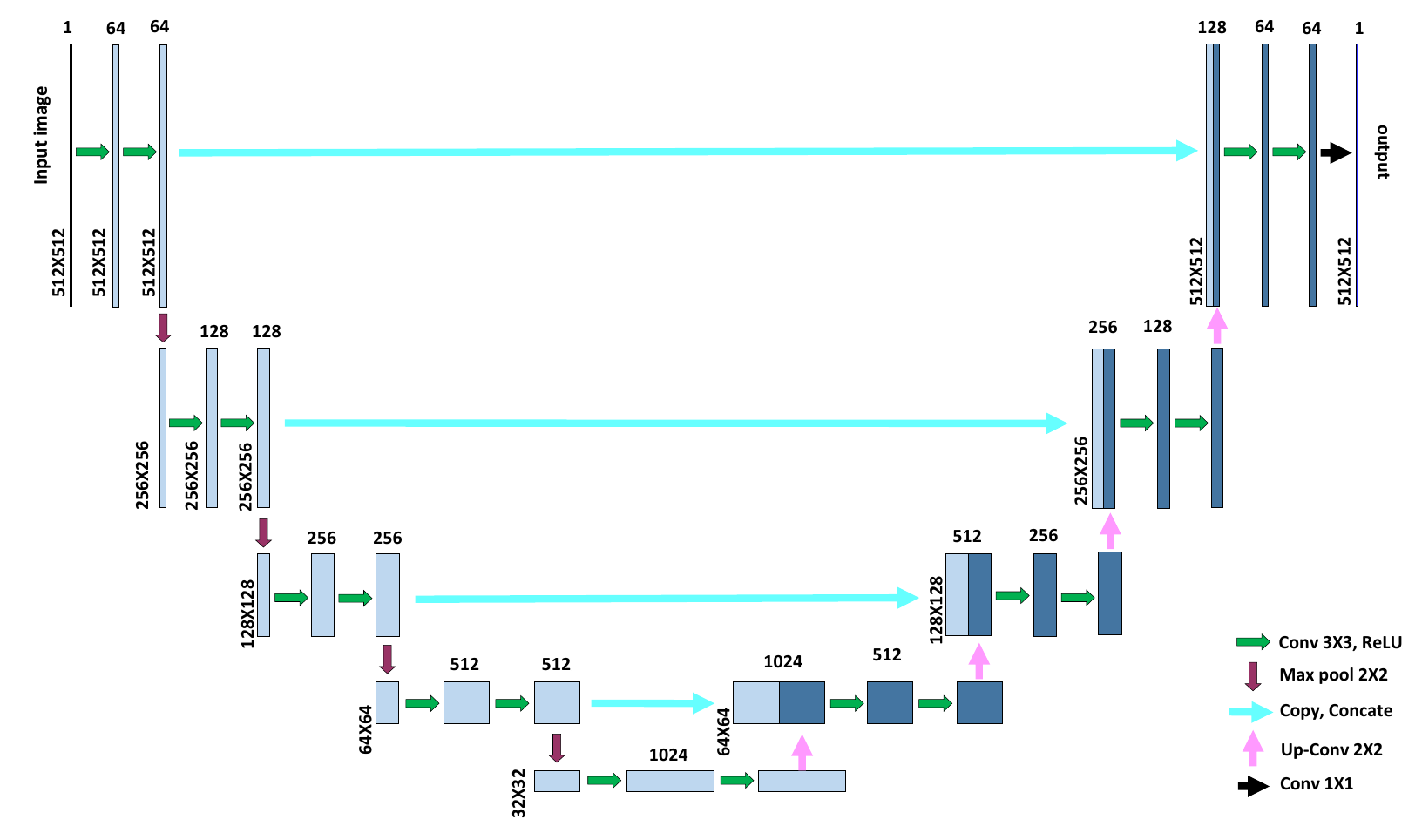}
\caption{U-Net.}
\label{fig:U-Net}
\end{figure}

In \cite{Ronneberger2015U-net}, a U-shaped deep neural network, called U-Net, was proposed for biomedical image segmentation which is by far one of the most successful deep image segmentation models. The architecture of U-Net is shown in Figure \ref{fig:U-Net}. It resembles the encoder/decoder architecture of AE if we view the left half of the U-Net as an encoder and the right half as a decoder. The main difference between the U-Net and AE is the use of skip connections of U-Net. Similar to the U-Net, \cite{milletari2016v} designed a DNN, called V-Net, for 3D volumetric medical image segmentation. Motivated by the U-Net and the convolutional framelets \cite{yin2017tale}, \cite{ye2018deep} designed a multi-resolution deep convolutional framelets. More recently, U-Net is extended to image analysis tasks \cite{falk2019unet}.

\section{Interpretations of Deep Neural Networks}\label{sec:interpretation_of_dl_architecture}

The development of traditional machine learning methods, such as support vector machine, decision tree, random forest, etc., benefit tremendously from theoretical studies in machine learning. However, existing machine learning theory, such as PAC, VC-dimension, Rademacher complexity, etc., may not be most suitable to analyze DNNs. Although DNNs are often composed of simple functions, such as convolutions, pooling, element-wise activation functions, etc., the entire networks are often difficult to analyze. Therefore, theoretical deep learning has become a popular area in machine learning that has attracted a lot of attention from theoretical computer scientists, statisticians, and mathematicians. In this section, we shall review some recent works on interpreting DNNs from two different perspectives, namely representation learning and differential equations. We will see that function approximation is a powerful tool in characterizing the efficacy of the given representation. It provides a rigorous analysis of the capacity of DNNs and how well they can approximate functions living in various function spaces. The perspective through differential equations, on the other hand, is more intuitive than function approximation and can explicitly guide the design of architectures of DNNs and training algorithms. There are also several other perspectives on the theoretical interpretations of deep learning. One may refer to the course ``Theories of Deep Learning" (STATS 385) hosted by David Donoho at Stanford University and the references therein (https://stats385.github.io/).

\subsection{Representation Learning Perspective}\label{sub:sparse_coding_perspective}

Images, such as medical images or natural images, are usually assumed to have sparse (or low dimensional) structures. The sparse structures can be effectively extracted by transformations. Successful examples include the (windowed) Fourier transform, wavelet transform, curvelet transform, etc, and they are able to provide efficient representations to images. They are pre-designed linear transformations and are independent of the given image data. DNNs can also be viewed as sparse representations that are able to extract sparse features from images. The difference is that DNNs are learned from a set of images and are (highly) nonlinear.

The quality of a given representation can be measured by its ability to approximate functions living in a certain function space. For example, let $\Phi:=\{\phi_i(\bm{x}):\bm{x}\in\mathbb{R}^n, i\in\mathbb{N}_+\}$ be a set of atoms, and function $f(\bm{x})$ be an element in function space $\mathcal{F}$ equipped with norm $\|\cdot\|$. One of the most basic and important approximation properties states as follows: for any given $\varepsilon>0$, there exists
$\tilde f_{\bm{\alpha},N}:=\sum_{i=1}^N\alpha_{i} \phi_{i}(\bm{x})$ with $N\in\mathbb{N}_+$ and $\bm{\alpha}=\{\alpha_1,\ldots,\alpha_N\}\in\mathbb{R}^{N}$ such that $$\|f-\tilde f_{\bm{\alpha},N}\|< \varepsilon.$$ A good representation requires fewer atoms (i.e. smaller $N$) to achieve an $\varepsilon$-approximation. The representation of various types of $\Phi$ have been well studied in the literature, such as polynomials, splines, Fourier basis and wavelets \cite{devore1993constructive,daubechies1992ten,Mallat2009}.

The neural network is a more efficient tool that can approximate a function arbitrarily well under suitable conditions \cite{hornik1991approximation,hornik1989multilayer,pinkus1999approximation}. Both the depth and width of a neural network are among the most important factors that affect its approximation power. In the following, we will review some of the existing characterizations of the approximation properties of shallow and deep neural networks.

Consider a shallow neural network with only one hidden layer
$$\tilde{f}_{N}(\bm{x};\bm\Theta)=\sum_{i=1}^{N}a_i\sigma(\bm{w}_{i}^\top\bm{x}+b_i),$$
where $\bm{x}\in\mathbb{R}^{n}$ is the input image data, $\bm{\Theta}=\{a_i,\bm{w}_{i},b_{i} \}, i=1,\ldots,N,$ are trainable parameters, and $\sigma(z)$ is an element-wisely applied nonlinear activation function. Examples of $ \sigma(z)$ are $\mbox{ReLU}(z)=\max(0,z)$, $\mbox{tanh}(z)=\frac{e^{z}-e^{-z}}{e^{z}+e^{-z}}$, $\mbox{sigmoid}(z)=\frac{1}{1+e^{-z}}$ and more generally a sigmoidal function \cite{cybenko1989approximation} that has the property:
\begin{equation}
\sigma(z)=
\begin{cases}
1 & \text{if}\; z\to +\infty,\\
0 & \text{if}\; z\to -\infty.
\end{cases}
\end{equation}
A DNN is a neural network with multiple hidden layers. It can be viewed as a successive composition of multiple shallow networks. A typical DNN (for regression problems) with depth $L$ and width $\bm N=(N_1,N_2,\ldots,N_L)$ denoted as
$$\tilde f_{L,\bm{N}}(\bm{x};\bm\Theta): \mathbb{R}^n\mapsto \mathbb{R},$$
can be recursively defined as: $\bm\Theta^\ell=(\bm\Theta^{\ell-1}, \theta^{\ell})$, $\tilde f_{\bm\Theta^\ell}=(\theta^\ell\circ\sigma\circ \tilde f_{\bm\Theta^{\ell-1}})$,  $\theta^\ell: \mathbb{R}^{N_\ell}\to \mathbb{R}^{N_{\ell+1}}$ with $\theta^{\ell}(\bm{x})=\bm{W}^\ell\bm x+\bm{b}^\ell$, and $\tilde f_{L,\bm{N}}:=\tilde f_{\bm\Theta^L}$.

Earlier results on the approximation property, i.e., universal approximation, suggest that a wide class of functions can indeed be approximated by neural networks with only one hidden layer, though the number of neurons, i.e. $N$, may increase exponentially as we decrease $\varepsilon$ \cite{funahashi1989approximate,barron1993universal,cybenko1989approximation}. There are benefits in increasing the depth $L$ of the neural network when approximating a target function. For example, approximation with DNNs leads to an exponential or polynomial reduction in the number of neurons while maintaining the same level of approximation accuracy \cite{liang2017deep,mhaskar2016learning,eldan2016power,cohen2016expressive}. \cite{delalleau2011shallow,telgarsky2015representation,telgarsky2016benefits} presented concrete examples that there exist functions that can be more efficiently represented with DNNs rather than shallow networks. In particular, \cite{telgarsky2016benefits} showed that the DNNs with $\mathcal{O}(L^3)$ layers and constant width cannot be approximated by networks with $\mathcal{O}(L)$ depth and less than $2^{L}$ width.

\cite{lu2017expressive} investigated the efficiency of depth of ReLU activated DNNs from a different angle by proving that there exist classes of wide neural networks which cannot be realized by any narrow network whose depth is no more than a polynomial bound. Comparing to the known result that there are classes of deep networks which cannot be realized by any shallow network whose size is no more than an exponential bound \cite{cohen2016expressive}, results from \cite{lu2017expressive} indicated that depth might be more effective than width. Although depth is more important than width, \cite{hanin2017approximating,hanin2017universal} proved that there is a minimum width of ReLU activated DNNs to ensure approximation of continuous functions. Their results indicated that a good DNN cannot be too narrow, otherwise we cannot approximate continuous functions even with infinite depth.

More recently, \cite{yarotsky2018optimal} analyzed the dependence of optimal approximation rate with respect to depth for ReLU activated DNNs. When approximating a multivariate polynomial, \cite{Rolnick2018the} proved that the total number of neurons in DNNs should grow linearly with respect to the number of variables of the polynomial.
\cite{shen2019nonlinear} provided intriguing analysis on ReLU activated DNNs via nonlinear approximation of composite dictionaries. They demonstrated the advantage of depth over width quantitatively by comparing the $N$-term approximation order of DNNs v.s. one-hidden-layer neural networks. Other than generic DNNs, theoretical analysis on the popular ResNet were also provided \cite{veit2016residual,lin2018resnet,e2019priori}.

In \cite{he2018relu}, the authors investigated the connection between linear finite element functions and ReLU deep neural networks. Firstly, they proposed an efficient ReLU activated DNN to represent any linear finite element functions and theoretically established that at least $2$ hidden layers are needed in a ReLU activated DNN to represent any linear finite element functions in
$\Omega \subseteq \mathbb{R}^d$ when $d\ge2$.
Then, using this relationship they established a straightforward error estimate as $\mathcal O(N^{-\frac{1}{d}})$ for a special kind of ReLU activated DNNs with $\mathcal {O}(N)$ non-zero parameters by involving the h-adaptive linear finite element approximation theory \cite{nochetto2011primer}.

Different from the approximation viewpoint, \cite{he2019mgnet} developed a unified model, known as MgNet,
that simultaneously recovers and extends some CNNs for image classification and multigrid methods for solving discretized PDEs,
by combining multigrid and deep learning methodologies.

\subsection{Differential Equation and Control Perspective}\label{sub:dynamics-perspective}

Given a DNN $\tilde f(\bm x;\bm\Theta): \mathbb{R}^n\to\mathbb{R}^m$, due to its composite structure as described in the previous subsection, we may view $\tilde f(\bm x;\bm\Theta)$ as an iterative mapping between $\mathbb{R}^n$ and $\mathbb{R}^m$. Then it is natural to view a generic DNN as a certain dynamic system \cite{cessac2010view}. However, a dynamic system that corresponds to a generic DNN is difficult to analyze since it does not have much special structure to exploit. Fortunately, it has been proven empirically that most of the effective DNNs have special structures in their architecture. In fact, designing special structures of DNNs, i.e. the architecture design, to make them easy to train and generalize better is one of the major research directions in deep learning. Furthermore, the objective of the emerging research topic neural architecture search (NAS), a subfield of automating machine learning (AutoML), is to search for effective DNN architecture for different data sets and tasks.

One of the most well-known DNNs with special structures is ResNet. Its bypasses (or shortcuts) enable us to efficiently train ultra-deep networks and achieve high accuracies in multiple tasks. The success of ResNet inspired the design of numerous new neural architectures. However, most of the design were based on empirical studies.
Although we can deploy NAS to search for new architectures, the current computation burden of NAS is still prohibitively high for researchers without access to heavy computation resources, and NAS cannot guarantee to find sufficiently new and interpretable neural structures. Therefore, we direly need a way to interpret ResNet and their siblings properly and to seek for guiding principles for the architecture design of DNNs.

Recently, \cite{weinan2017proposal} made an inspiring observation that ResNet can be viewed as forward Euler scheme solving for an ordinary differential equation (ODE), and links training of DNNs with optimal control. \cite{sonoda2017double} and \cite{li2017deep} also regarded ResNet as dynamic systems that are the characteristic lines of a transport equation on the data distribution. Similar observations were also made by \cite{chang2018reversible,chang2017multi}. A rigorous justification of the link between ResNet and ODEs was given by \cite{thorpe2018deep}, and that of the link between deep learning and optimal control was given by \cite{weinan2019mean}. The dynamics and control perspective enabled us to design more efficient algorithms solving related deep learning problems \cite{li2017maximum,chen2018neural,zhang2018dynamically,zhang2019you}.

In \cite{lu18beyond}, the authors suggested a general bridge between numerical ODEs and deep neural architectures by observing that multiple state-of-the-art deep network architectures, such as PolyNet \cite{zhang2017polynet}, FractalNet \cite{Larsson2016FractalNet} and RevNet \cite{gomez2017reversible}, can be viewed as different discretizations of ODEs. Furthermore, \cite{lu18beyond} proved that ResNet with certain stochastic training strategies weakly approximates stochastic differential equations, which granted stochastic control perspective on randomized training of DNNs. More importantly, such new perspectives enable us to systematically design deep neural architectures through numerical (stochastic) differential equations, which is a rather mature field in applied mathematics. In this section, we shall review some of the findings of \cite{lu18beyond} and some other related works.

\subsubsection{Numerical difference equations and architecture design}

\begin{figure}[h]
\centering
\subfigure[ResNet]{\includegraphics[scale=0.3]{ResNet}\label{ResNet}}
\subfigure[PolyNet]{\includegraphics[scale=0.36]{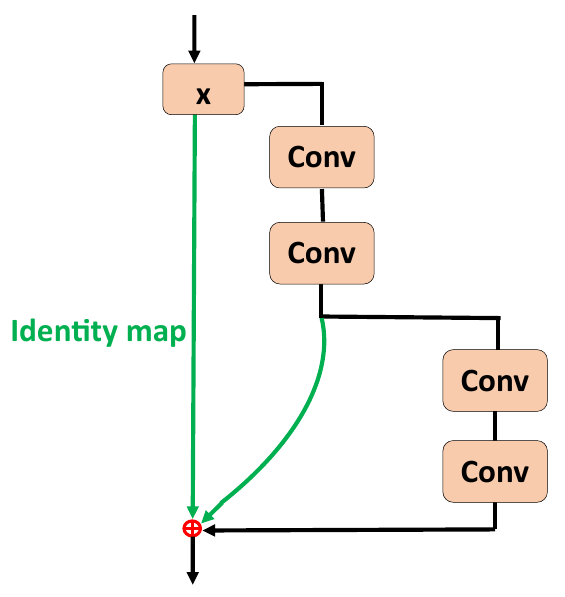}\label{PolyNet}}
\subfigure[FractalNet]{\includegraphics[scale=0.36]{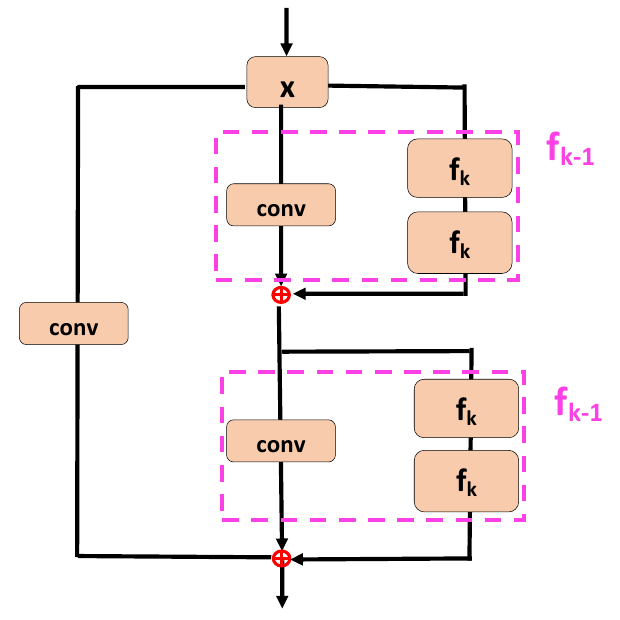}\label{FractalNet}}\\
\subfigure[RevNet]{\includegraphics[scale=0.38]{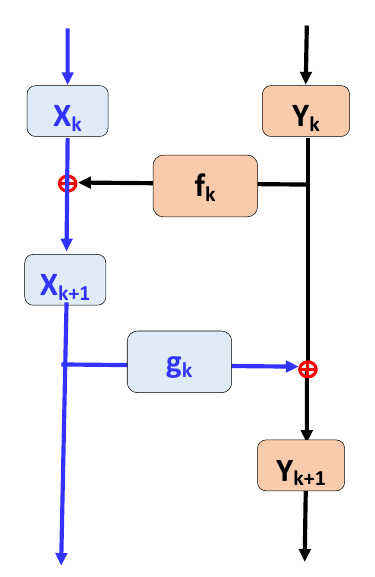}\label{RevNet}}
\subfigure[LM-ResNet]{\includegraphics[scale=0.3]{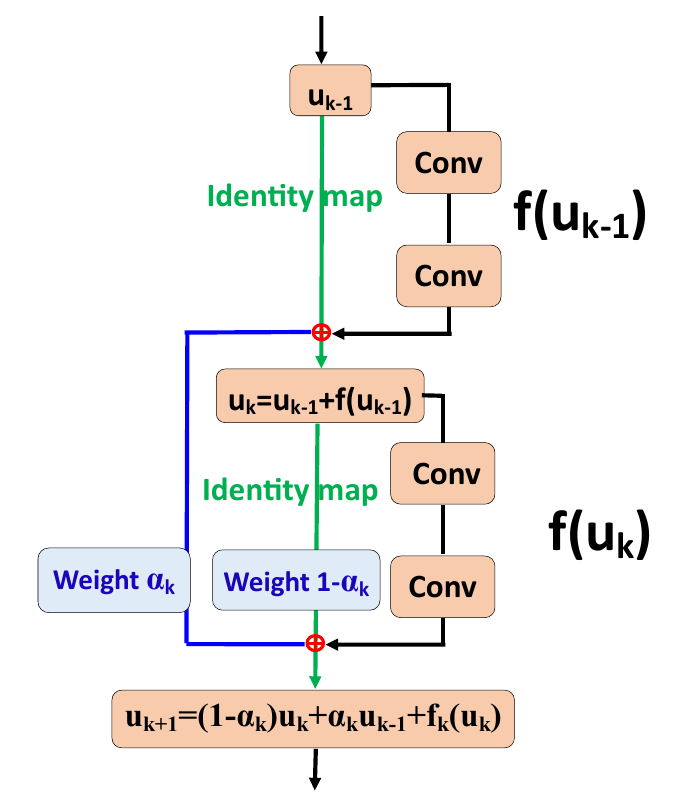}\label{LM_structure}}
\caption{
Schematics of different neural network architectures
}
\label{fig:LM-structure}
\end{figure}

We first show that how ResNet is related to forward Euler scheme in numerical ODEs. Considering a building block of ResNet \eqref{eq:resnet_block} as shown in Figure\ref{fig:ResNet}, it can be rewritten as
$$\uu_{k+1}=\uu_{k}+\Delta t_k \FF(\uu_k, t_k),$$
or equivalently as
$$\frac{\uu_{k+1}-\uu_{k}}{\Delta t_k}=\FF(\uu_k,t_k),$$
where $ \Delta t_k$ is the step size and $ \FF (\uu_k,t_k)= \frac{1}{\Delta t_k}\FF_{k}(u_k)$. The above formula is the forward Euler scheme solving the following ordinary differential equation (ODE)
\begin{equation}\label{eq:resnet_ode}
\frac{d\uu}{dt}=\FF(\uu,t).
\end{equation}
Therefore, the ResNet can be viewed as the forward Euler discretization of the ODE \eqref{eq:resnet_ode} with step size $\Delta t_k=1$ for every $k$. This was first observed by \cite{weinan2017proposal}. More recently, \cite{zhang2019towards} showed that there are benefits of considering ResNet with $0<\Delta t_k<1$.

In \cite{lu18beyond}, the authors further observed that many other DNNs with bypasses, e.g., PolyNet \cite{zhang2017polynet} (Figure \ref{PolyNet}), FractalNet  \cite{Larsson2016FractalNet} (Figure \ref{FractalNet}) and RevNet  \cite{gomez2017reversible} (Figure \ref{RevNet}), can be interpreted as certain temporal discretizations of ODEs. For example, the PolyInception module (Figure \ref{PolyNet}) of PolyNet can be written mathematically as
$$(I+\FF+\FF^2)(x)=x+\FF(x)+\FF(\FF(x)),$$
where $I$ is the identity map, $\FF$ is a nonlinear operator and $x$ is the input feature map. Note that the above polynomial of mapping $\FF$ is an approximation of $(I-\FF)^{-1}$ using a truncated Neumann series: $$(I-\varDelta t \FF)^{-1}\approx (I+\varDelta t\FF+\varDelta t^2\FF^2).$$
Therefore, PolyNet can be viewed as an approximation to the backward Euler scheme solving the ODE \eqref{eq:resnet_ode}. FractalNet (Figure \ref{FractalNet}) can be viewed as approximation of the ODE \eqref{eq:resnet_ode} with Runge-Kutta scheme. See \cite{lu18beyond} for more examples and further details.

These examples suggest a potential link between numerical ODEs and deep neural architecture. A remaining question is whether deep neural architecture design can benefit from such perspective. The authors of \cite{lu18beyond} designed a new ResNet-like module, called the linear multi-step structure (LM-structure) using the linear multi-step schemes in numerical ODEs \cite{ascher1998computer}. The LM-structure (linear two-step structure to be more precise) can be written mathematically as
\begin{equation}\label{eq:LM-structure}
\uu_{k+1}=(1-\gamma_{k})\uu_{k}+\gamma_{k}\uu_{k-1}+\FF(\uu_{k},t_k),
\end{equation}
where $\gamma_{k}\in \mathbb{R}$ is a trainable parameter in each layer. Note that when $\gamma_k=0$ for all $k$, the LM-structure reduces to ResNet. Figure \ref{LM_structure} shows the LM-structure. Empirical results of \cite{lu18beyond} showed that the LM-structure boost classification accuracies of ResNet-like DNNs on CIFAR and ImageNet. It can also reduce the depth (hence number of parameters) of ResNet-like DNNs by 50--90\% without hampering accuracies. Other than the LM-structure, one can use the mid-point scheme or the leapfrog scheme to design new DNNs \cite{chang2018reversible}, or using the Runge-Kutta method \cite{zhu2018convolutional}.

The performance gain of the LM-structure can be explained using the concept of modified equations \cite{warming1974modified}.
By Taylor's expansion, the modified equation associated with the LM-structure \eqref{eq:LM-structure} is
\begin{equation}
(1+\gamma_{k})\dot{\uu}_{k}+\frac{1-\gamma_{k}}{2}\Delta t \ddot{\uu}_{k}
=\FF(\uu_{k},t).
\end{equation}
Comparing to ResNet, the LM-structure has the freedom to balance between $\ddot{\uu}_{k}$ of $\uu_{k}$. Having bigger weights on $\ddot{\uu}_{k}$ can speed up the information propagation of the dynamics as shown by various earlier work such as \cite{su2014differential,wilson2016lyapunov,dong2017image}. This is why LM-structure can achieve comparable accuracies with a much smaller depth than ResNet and its siblings.

\subsubsection{Stochastic training and optimal control}

Stochastic training, such as dropout and noise injections, is widely adopted in training of DNNs. It helps with the generalization of the trained networks. In \cite{lu18beyond}, the authors showed that some stochastic training of ResNet, shake-shake \cite{gastaldi2017shake} and stochastic depth \cite{huang2016deep}, can be viewed as stochastic control
\begin{eqnarray}\label{eq:stochastic:control}
\min_{\bm\Theta} \mathbb{E}_{\bm X(0)\sim data}\left[ \mathbb{E}[L(\bm X(T))]+\int_{0}^{T} R(\bm\Theta)\right]\notag\\
s.t. \quad d\bm X=\FF(\bm X,\theta)dt+\mathcal G(\bm X,\theta)d\bm B_{t},
\end{eqnarray}
where the stochastic differential equation \eqref{eq:stochastic:control} is the weak limit of the ResNet with shake-shake mechanism or stochastic depth. This suggests a connection between stochastic training and stochastic control, and a connection between DNNs with randomness and stochastic differential equations. Later, \cite{sun2018stochastic} observed that the stochastic training of ResNet and its variants is closely related to the optimal control of backward Kolmogorov's equations, and the popular dropout regularization essentially introduces viscosity to the equations.

\section{Deep Models in Medical Image Reconstruction}\label{sec:imaging_models}

Classical medical image reconstruction methods, such as FBP and algebraic reconstruction method (ART) for CT imaging, are highly efficient and widely used in practice \cite{natterer2001mathematics}. However, these methods are also prone to be sensitive to noise and incompleteness of measured data. To obtain a high-quality image, numerous regularization based models and algorithms have been developed \cite{zeng2010medical,Scherzer2015Handbook,herman2009fundamentals} in the past three decades. In recent years, there has been a continuous effort in the medical imaging community to further advance medical image reconstruction by combining traditional image reconstruction methods with deep learning. When combining traditional handcraft modeling with deep modeling, two general approaches are often adopted: post-processing and raw-to-image. The validity of these two approaches are generally supported by, though still rather incomplete, the analysis on the approximation properties of DNNs as described in Section \ref{sub:sparse_coding_perspective}, and by the dynamics perspective on the DNNs with certain special structures as described in Section \ref{sub:dynamics-perspective}.

For post-processing, one needs to estimate the mapping between the initially reconstructed low-quality image and its high-quality counterpart. This is possible since DNNs can approximate generic functions or mappings as discussed in Section \ref{sub:sparse_coding_perspective}. This approach is effective mostly when the initial reconstruction and its high-quality counterpart are not drastically different. However, due to limited measurements and the presence of noise, the initially reconstructed image may contain heavy and complex artifacts which are difficult to remove even by deep models. Furthermore, the information missing from the initial reconstruction cannot be reliably recovered by any post-processing. Thus, the post-processing approach has limited performance and is more suitable to handle initial reconstructions that are of relatively high quality.

For raw-to-image, one directly estimates the mapping between the raw data (e.g. the projection data of CT and k-space data for MRI) and the reconstruction image. The challenge, however, is that the data distribution in the domain of raw data is often vastly different from that in the image domain. Learning a direct mapping using a DNN without special structures (e.g., a fully connected network or a vanilla CNN), though not impossible, may require tons of training data, can be computationally expensive and heavily relies on good initializations of the model parameters (e.g., the AUTOMAP \cite{zhu2018image}). It is well-known in the literature of handcrafted modeling that the mapping ought to have certain dynamic structures which can be represented by a carefully designed (partial) differential equation or an optimization algorithm solving a certain objective function(al). Thus, it is more plausible to combine handcrafted dynamics with deep learning. The way of such combination was depicted in Section \ref{sub:dynamics-perspective} in a relatively general setting where we did not discuss how $\mathcal{F}$ should be designed for a given image restoration problem. Nonetheless, it is rather convincing that there are connections between dynamic systems and DNNs and the benefits of recognizing such connections.

Our rich history of handcraft modeling in image restoration provides us with an abundant set of tools that we can select freely for the mapping $\mathcal{F}$ via the general technique known as the unrolling dynamics \cite{gregor2010learning,sun2016deep}. To be more precise, this approach first suggests us to unroll optimization algorithms that are introduced to solve handcrafted models into feed-forward networks. Then, we incorporate our domain knowledge of the problem in-hand to determine which parameters are best to be learned from the data in an end-to-end fashion. The advantage of designing deep models via unrolling optimization algorithms is threefold: 1) the deep model through unrolling dynamics is more interpretable than a regular deep model such as U-Net; 2) the number of parameters are normally less than regular deep models and thus more suitable for small sample learning; 3) it provides a general way of combining domain knowledge with deep learning so that we can easily decide on which component in the model need to be learned and which can be handcrafted without losing much expressive power of the model.

As mentioned in the introduction that one of the major differences between medical image reconstruction and image restoration in computer vision is the quality metric of the reconstruction images. It has long been discussed in the medical imaging community that such a quality metric is best, in many scenarios, to be task-based rather than generic metrics such as PSNR and SSIM. The importance of providing such a task-based metric for medical imaging was recently discussed in the article \cite{kalra2018radiomics}. The question is, however, how can we realize such task-based image quality metric? Recently, the authors of \cite{wu2017end} proposes to realize task-based quality metric by ``hooking" a image reconstruction network from unrolling dynamics with a image analysis DNN, so that the reconstructed images by the first network will be implicitly evaluated by the second which effectively makes the quality metric task-based. Similar idea appeared in computer vision for image denoising \cite{liu2017image,liu2018connecting}. On the other than, these work also suggested a new ``raw-to-task" modeling philosophy with encouraging empirical results. Therefore, the entire pipeline of image reconstruction, analysis, and decision making can be effectively integrated.

In the rest of this section, we provide more details on the aforementioned models.

\subsection{Post-Processing}\label{sub:post_processing}

\begin{figure}[h]
\centering
\subfigure[Ground truth]{\includegraphics[scale=0.43]{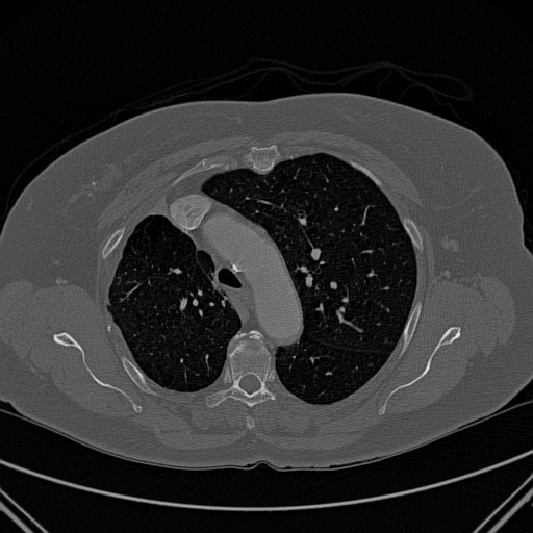}\label{ground_truth}}
\subfigure[Low dose CT ]{\includegraphics[scale=0.43]{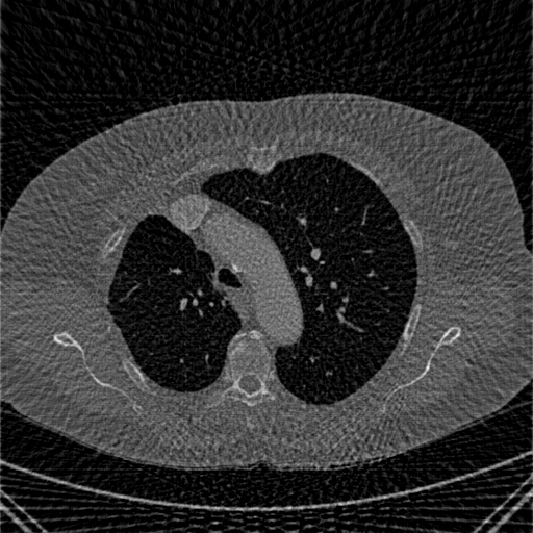}\label{low_dose}}
\subfigure[Low tube current CT ]{\includegraphics[scale=0.43]{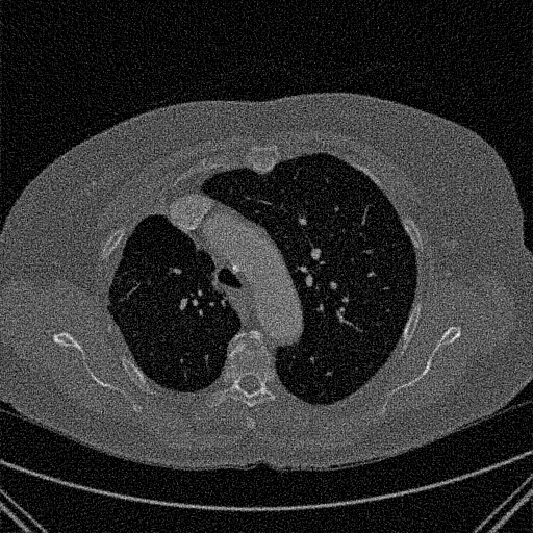}\label{low_tube_current}}
\caption{
FBP reconstructed images
}
\label{fig:low-dose-noisy-CT}
\end{figure}

Post-processing is a procedure to enhance the quality of an initially reconstructed image. In this subsection, we use CT as an example. Due to the incompleteness of measured data in sparse view and limited angle CT, the FBP reconstructed image is often degraded by streaking artifacts (Figure \ref{low_dose}). Noise caused by low tube currents is another source of degradation of CT images (Figure \ref{low_tube_current}).
In \cite{chen2017low,zhang2018sparse}, a residual encoder-decoder CNN (Figure \ref{fig:autoencoder}) was used to approximate the map between the degraded image and the clean image. This model is efficient in removing noise from the FBP reconstructed CT images. To protect subtle structures in CT images while suppressing noise, \cite{yang2018low} adopted a generative adversarial network (GAN) with the loss function defined by a combination of Wasserstein distance and the perceptual difference between input degraded image and the corresponding clean image.

To reduce the radiation dose and acquisition time, one can decrease the number of projections of X-ray CT, which is known as the sparse view or limited angel CT. Such incompleteness of measurements leads to streaking artifacts with global and yet relatively simple structures in the FBP reconstructed CT images. In this case, a DNN with multi-scale architecture can be used to capture the global patterns of streaking artifacts. With such observation, \cite{Jin2017Deep,han2016deep} utilized the U-Net (Figure \ref{fig:U-Net}) to reduce artifacts in FBP reconstructed sparse view CT images. The repaired high-quality CT image is the subtraction of the learned artifacts by the U-Net from the degraded input image. In some sense, U-Net takes a role of residual learning \cite{He2016Deep}.

\subsection{Raw-to-Image}\label{sec:deep_model_design}

We describe how optimization algorithms can be unrolled and set up as a deep feed-forward network for end-to-end training. We remark that, under some specific conditions, the learning empowered optimization algorithms via unrolling dynamics can have better provable convergence than the original optimization algorithms \cite{chen2018theoretical,liu2019alista,xie2019differentiable}. This was in fact the original motivation of \cite{gregor2010learning} to use machine learning to improve  optimization algorithms. In this subsection, however, we shall focus on the ``dual" aspect of unrolling dynamics, i.e. how optimization algorithms inspire new and more effective deep network architectures for medical image reconstruction or inverse problems in general.

\subsubsection*{ADMM-Net}

The work of ADMM-Net proposed by \cite{sun2016deep} was the first to suggest the potential benefit of designing deep neural networks for inverse problems by unrolling optimization algorithms.

In the iteration scheme \eqref{eq:ADMM-scaled-form} of ADMM algorithm (Subsection \ref{ssub:admm_algorithm}),
the tuning parameters such as $\mu,\lambda $ and $\beta_{k}$, the handcrafted operator $\WW$ and function $\Phi$ are difficult to determine adaptively for a given data set. In \cite{sun2016deep}, the authors proposed to unroll the ADMM algorithm to design a new deep model, named ADMM-Net. By doing so, the tuning parameters and the predefined linear operator $\WW$ are now all learnable from the training data. The proximal operator of the sparsity promoting function $\Phi$ is parameterized by a piecewise linear function with learnable parameters as well. As a result, the thresholding operator $\TT_{\lambda}(\cdot)$ in ADMM algorithm is also learned from the training data.
In a basic version of ADMM-Net \cite{sun2016deep}, $\dd^{k+1}$ is updated by
\begin{equation}
\dd^{k+1}=\TT_{\bm{\varTheta}_{1}}\left(\calW_{\bm{\varTheta}_{2}}(\uu^{k+1})+\bb^{k}\right),
\end{equation}
where $\TT_{\bm{\varTheta}_{1}}(\cdot)$ is a parameterized piecewise linear function with parameters $\bm{\varTheta}_{1}$, and $\calW_{\bm{\varTheta}_{2}}$ is a parameterized convolution layer with parameters $\bm{\varTheta}_{2}$. The ADMM-Net was later further improved by \cite{yang2017admm} and the new model was called the Generic-ADMM-Net where different variable splitting strategy was adopted in the derivation of the ADMM algorithm. The Generic-ADMM-Net achieved state-of-the-art MR image reconstruction results with a significant margin over the BM3D-based algorithm.

\subsubsection*{Primal-Dual Networks (PD-Net)}

In \cite{adler2018learned}, the authors unrolled the iteration scheme \eqref{eq:PD-resolvent} and \eqref{eq:PD-error-correct} of the PDHG algorithm to design new deep model for CT image reconstruction. This new deep model was called the primal-dual network (PD-Net).
The main idea is to approximate each resolvent/proximal operator  \cite{parikh2014proximal} in the subproblem of PDHG by a neural network. Thus, it circumvents the difficulties in choosing optimal forms $\Phi$ and $F$. One layer of PD-Net takes the form
\begin{eqnarray}\label{eq:PD-Net}
\ww^{k+1}&=&\calN_{\ww}\left([\ww^{k},\Wu^{k}]; \bm{\Theta}_{\ww}^{k}\right),\notag\\
\uu^{k+1}&=&\calN_{\uu}\left([\uu^{k},\WW^{\top}\ww^{k+1},\bm{A}^{\top}\ff]; \bm{\Theta}_{w}^{k}\right),
\end{eqnarray}
where $\bm{f}$ is the measured data, $\bm{A}$ is the imaging operator, $\calN_{\ww}(\cdot;\bm{\Theta}_{\ww}^{k} )$ and $\calN_{\uu}(\cdot;\bm{\Theta}_{\uu}^{k} )$ are neural networks parameterized by $\bm{\Theta}_{\ww}^{k} $ and $\bm{\Theta}_{\uu}^{k}$ respectively. The notation $[\bm{v}_1,\cdots,\bm{v}_m]$ denotes concatenation of the components $\bm{v}_1, \ldots, \bm{v}_m$ into a higher dimension tensor.
The linear operator $\WW$ can be either fixed or learned from the data. PD-Net has a significant performance boost compared with FBP and some handcrafted reconstruction models \cite{adler2018learned,Adler2017Solving}.

\subsubsection*{JSR-Net}

To suppress the artifacts induced by incomplete data and noise, \cite{Dong2013} proposed a joint spatial-Radon domain reconstruction (JSR) model for sparse view CT imaging as following
\begin{equation}\label{JSR-classical}
\min_{\uu,\ff} \FF(\uu,\ff,\YY)+\mathcal{R}(\uu,\ff),
\end{equation}
where the data fidelity term $\FF(\uu,\ff,\YY)$ is defined by
$$\FF(\uu,\ff,\YY)=\frac{1}{2}\|R_{\bm{\Gamma}^{c}}(\ff-\YY)\|^{2}+\frac{\alpha}{2}\|R_{\bm{\Gamma}}(\bm{A}\uu-\ff)\|^{2}+\frac{\gamma}{2}\|R_{\bm{\Gamma}^{c}}(\bm{A}\uu-\YY)\|^{2},$$
and the regularization term defined by
$$\mathcal{R}(\uu,\ff)=\|\bm{\lambda}_{1}\cdot \WW_{1}\uu\|_{1,2}+\|\bm{\lambda}_{2}\cdot\bm{W}_{2}\ff\|_{1,2}.$$
The notation $ R_{\bm{\Gamma}}$ is a restriction operator with respect to the missing data region indexed by $\bm{\Gamma}$. $ R_{\bm{\Gamma}}$ takes value $1$ if the element's index contained in $\bm{\Gamma}$ and $0$ elsewhere. Here, $\bm{\Gamma}^{c}$ indicates the region of available measured data and is the complement of $\bm{\Gamma}$.  $\bm{A}$ is a discrete form of the Radon transform, $\YY$ is the measured projection data. Note that, in JSR model $\uu$ and $\ff$ are the underlying CT image and the restored high-quality projection data respectively. $\WW_{i},i=1,2$, are tight frame transforms and $\bm{\lambda}_{i},i=1,2$, are the regularization parameters.

The handcrafted JSR model \eqref{JSR-classical} enforces the data consistency in the Radon domain and image domain simultaneously. Thus, it leads to improved quality of the reconstructed image. Similar data fidelity design was adopted in \cite{burger2014total} to model the positron emission tomography. Later, \cite{zhan2016ct} propose to improve the JSR model by learning the tight frame transforms $\WW_i$ from the data. More recently, a re-weighting strategy was introduced in JSR model to reduce the metal artifacts in multi-chromatic energy CT  \cite{zhang2018reweighted}.

Existing work showed the potential of the JSR framework, and it is natural to consider using unrolling to derive a deep model from algorithms solving the JSR model. In \cite{zhang2019JRSNet}, the authors designed the JSR-Net for sparse view and limited angle CT image reconstruction. The JSR-Net is derived by unrolling an alternative optimization algorithm with subproblems solved by ADMM. Similar to the PD-Net, JSR-Net also adopted neural networks to approximate the proximal operators. The advantage of JSR-Net is that it can efficiently utilize multi-domain image features to improve the quality of the reconstructed image.

\subsection{Raw-to-Task}\label{sec:task_driven}

\begin{figure}[h]
\centering
\subfigure[Image reconstruction ]{\includegraphics[scale=0.4]{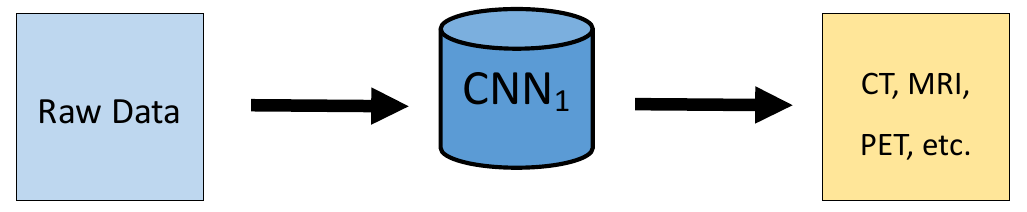}\label{task_driven_workflow_rec}}\\
\subfigure[Image analysis]{\includegraphics[scale=0.4]{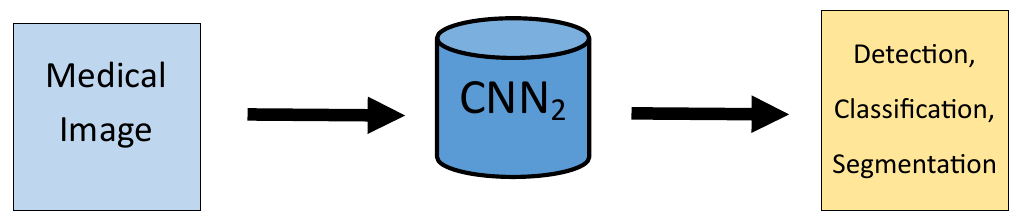}\label{task_driven_workflow_analysis}}\\
\subfigure[Image reconstruction and analysis]{\includegraphics[scale=0.4]{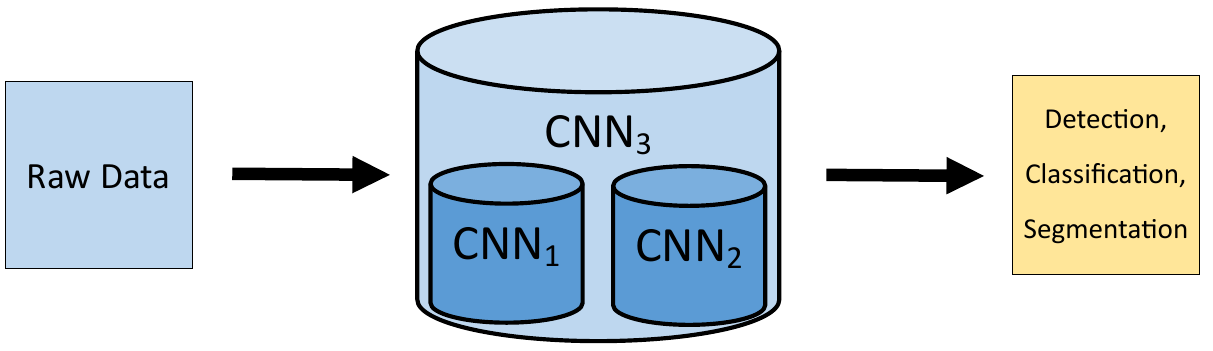}\label{task_driven_workflow_rec_analysis}}
\caption{
CNN based workflows for medical image reconstruction and analysis.
}
\label{fig:task-driven-workflow}
\end{figure}

The traditional workflow of medical image analysis has two separate stages: 1) reconstruction of a high-quality image from raw data (see Figure \ref{task_driven_workflow_rec}), and 2) make a diagnosis based on the high-quality reconstructed image (see Figure \ref{task_driven_workflow_analysis}). The drawbacks of the two-stages' approach and the potential benefit of uniting the two stages were discussed earlier. Here, we shall describe how we can join the two stages into one unified step (see Figure \ref{task_driven_workflow_rec_analysis}).

As discussed in Section \ref{sec:deep_model_design} that we can design feed-forward deep networks for image reconstruction. Once we have an image, there are plenty of choices of deep neural networks for various image analysis tasks. The most simple and natural way of joining image reconstruction and image analysis is to connect the two networks together and conduct end-to-end training (from scratch or by fine-tuning). Such idea was first introduced by \cite{wu2017end} in medical imaging and by \cite{liu2017image,liu2018connecting} in computer vision for image denoising. By doing so, the second network for image analysis can be regarded as a task-based image quality metric that is learned from the data. As shown in \cite{wu2017end}, where the image analysis task was lung nodule recognition, the learned image quality metric automatically placed more emphasis within the lung areas and less emphasis elsewhere. Such a quality metric is specific to the task of lung nodule recognition since the image quality outside of the lung region is irrelevant to the task.

\section{Challenge and Opportunities}\label{sec:challenge_and_opportunities_in_deep_learning}

Although deep learning based models continue to dominant medical imaging, there are still plenty of remaining challenges in deep modeling which limit the application and implementation of these new methods in clinical practice. These challenges also present themselves as new opportunities for researchers working in related fields.

\begin{itemize}
    \item The everlasting hunger of labeled data. There are only limited labeled data available to develop new deep models in medical imaging. Annotation of medical images is time-consuming and requires expert knowledge from physicians. Can we design effective learning models that can make good use of both the (very limited) labeled data and the (relatively more abundant) unlabelled data?
    \item The limited number of observations. Due to morbidity and privacy concerns, it is generally difficult to gather very large medical data for a specific task. Furthermore, the number of rare cases is (by definition) small but can be much more valuable than common cases. Can we design learning models and data augmentation techniques to effectively extract knowledge from these limited samples and acknowledge such unequal importance among the samples?
    \item Radiologists do not make the clinical decision only based on images. More information from the patients and the knowledge of the doctors from their years of training in medical school are also crucial in decision making. Thus, incorporating data gathered from multiple diverse sources into deep modeling is important in improving system performance.
    \item Reasoning is just as important as, if not more important than, inferencing. Currently, most deep models hide the reasoning procedure. There is a chance that the model makes accurate predictions based on wrong reasoning. This makes the model unreliable. Can we incorporate deep modeling with reasoning (such as causal inference) or with medical knowledge graph? This may further reduce the amount of annotated data we need to train deep models without hurting performance.
\end{itemize}

\end{document}